\documentclass[amssymb,useAMS,prd,aps,amsmath,twocolumn,superscriptaddress,nofootinbib,preprintnumbers]{revtex4}
\pdfoutput=1
\usepackage{graphicx}
\usepackage{psfrag}
\usepackage{braket}
\usepackage[usenames,dvipsnames]{color}
\usepackage[colorlinks=true, linkcolor=BrickRed, citecolor=Blue, urlcolor=Blue, filecolor=Blue, draft]{hyperref}
\usepackage{epsfig,amsmath,amssymb,fancyheadings}   
\usepackage[normalem]{ulem}

\newcommand{\beq}{\begin{equation}}
\newcommand{\eeq}{\end{equation}}
\newcommand{\ben}{\begin{eqnarray}}
\newcommand{\een}{\end{eqnarray}}
\newcommand{\bi}{\begin{itemize}}
\newcommand{\ei}{\end{itemize}}

\newcommand{\remove}[1]{}

\def\refe@jnl#1{{#1}}
\def\aj{\refe@jnl{Astron.~J.}}
\def\araa{\refe@jnl{Annu.~Rev.~Astron.~Astrophys.}}
\def\apj{\refe@jnl{Astrophys.~J.}}
\def\apjl{\refe@jnl{Astrophys.~J.~Lett.}}
\def\aap{\refe@jnl{Astron.~Astrophys.}}
\def\mnras{\refe@jnl{Mon.~Not.~R.~Astron.~Soc.}}
\def\prd{\refe@jnl{Phys.~Rev.~D}}
\def\fcp{\refe@jnl{Fund.~Cos.~Phys.}}
\def\physrep{\refe@jnl{Phys.~Rep.}}
\def\physlett{\refe@jnl{Phys.~Lett.}}

\def\invisible#1{  }

\def\pe{{p_{e^-}}}

\def\pe1{{p_{e_1}}}
\def\ghost#1{{ (here is the text made invisible)}}

\begin{document}

\title{ Circular polarisation: a new probe of dark matter and neutrinos in the sky}  

\author{C\'eline B\oe hm} 
\email{c.m.boehm@durham.ac.uk}
\affiliation{Institute for Particle Physics Phenomenology, Durham University, South Road, Durham, DH1 3LE, United Kingdom}
\affiliation{LAPTH, U. de Savoie, CNRS,  BP 110, 74941 Annecy-Le-Vieux, France}
\affiliation{Excellence Cluster Universe, Boltzmannstr. 2, D-85748, Garching, Germany}

\author{C\'eline Degrande}
\affiliation{CERN, Theory Division, Geneva 23 CH-1211, Switzerland}

\author{Olivier Mattelaer}
\affiliation{Center for Cosmology, Particle Physics and Phenomenology, Universit\'e Catholique de Louvain 
Chemin du cyclotron,2 1348 Louvain-La-Neuve Belgium}

\author{Aaron C. Vincent}
\affiliation{Department of Physics, Imperial College London, Blackett Laboratory, Prince Consort Road SW7 2AZ, United Kingdom}

\preprint{CERN-TH-2017-003,CP3-17-01}
\begin{abstract}
The study of anomalous electromagnetic emission in the sky is the basis of indirect searches for dark matter. It is also a powerful tool to constrain the radiative decay of active neutrinos. Until now, quantitative analyses have  focused on the flux and energy spectrum of such an emission; polarisation has never been considered. Here we show that we could be missing out on an essential piece of information. The radiative decay of neutrinos, as well as the interactions of dark matter and neutrinos with Standard Model particles can generate a  circular polarisation signal in X-rays or $\gamma$-rays. If observed, this could reveal important information about their spatial distribution and particle-antiparticle ratio, and could even reveal the nature of  the high-energy particle physics processes taking place in astrophysical sites. The question of the observability of these polarised signatures and their separation from background astrophysical sources is left for future work.

\end{abstract}
\maketitle

\section{Introduction}
Nonzero neutrino masses and the existence of dark matter are two of the most robust indications that the Standard Model (SM) of particle physics is incomplete. Unsurprisingly,  the ``dark sector'' particles are the least well known constituents of the Universe.  
We do not know the absolute masses of the active neutrinos, nor do we know the mechanism by which they acquire a mass.  
More puzzling even, we do not know if the dark matter is made of particles \cite{Bird:2016dcv,Carr:2016drx}.

In the past, observations of the electromagnetic emission from supernovae, the sun, and the diffuse photon spectrum have been used to place limits the neutrino decay rate \cite{Raffelt:1985rj,Raffelt:1985nm,Ressell:1989rz,Kainulainen:2002pu}.  Extragalactic and galactic electromagnetic emission  in the sky (from clusters of galaxies, galaxies and dwarf galaxies) has also been studied in the hope of unveiling signatures of dark matter annihilation or decay (see \cite{Gunn:1978gr,Stecker:1978du} for historical references). No signal has been found as yet. 

A major challenge for indirect searches is to disentangle a  new physics signature from the numerous Standard Model sources of photons in the sky (which includes stars, jets, supernovae and the big bang itself). Exploiting the polarisation of the signal could be the  way forward, but one first needs to determine which processes lead to a linear and circular polarisation at high energy 
(X and $\gamma$-rays). 

$\gamma$-ray circular polarization can be measured thanks to the correlation with the spin of an outgoing electron after Compton scattering, measurable via its Bremsstrahlung signature \cite{Tashenov2011164,Tashenov2014}. $\gamma$-ray polarimetry has in fact been used for decades to search for CP-violating processes in nature, and was key in determining the helicity of the neutrino \cite{PhysRev.109.1015}.

Circular polarisation is a key phenomenon which has been extensively studied in astrophysics but has not been exploited in astroparticle physics. 
In astrophysics, the circular polarisation  of the radio spectrum  is a clear indication of synchrotron emission and a powerful probe of the presence of high energy electrons  \cite{Longair:1992ze,deBurca:2015kea}.

In Ref.~\cite{Gorbunov:2016zxf} it was suggested on the basis of symmetry considerations that the decay of asymmetric dark matter particles (such as sterile neutrinos) could induce circularly polarised X-ray line signals; Ref \cite{Kumar:2016cum} showed that under certain conditions, dark matter annihilation via a charged mediator could give a net polarized state.  Here, we show that the observation of circularly polarised X or $\gamma$-rays could reveal even more. 

The fraction of polarisation and the energy dependence of the polarised spectrum could be used to probe the existence of neutrino-electron interactions in astrophysical sites, as well as reveal dark matter interactions with ambient cosmic rays. 
If the signal can be separated from the synchrotron background, then circular polarisation signals could be used to track the dark matter and cosmic rays spatial  distribution.  
 
In Section  \ref{sec:recap}, we recall the conditions for circular polarisation to be produced. In  Section  \ref{sec:neutrinos} and \ref{sec:DM}, we examine examples of neutrino and DM processes that can give rise to circular polarisation. We will present many examples, but we stress that the list of processes studied here is not exhaustive.  We conclude in Section  \ref{sec:conclusion}.

\section{ Generation of circular polarisation \label{sec:recap}}

\subsection{Polarisation vectors}
In Quantum mechanics, a photon with momentum $k^\mu=(k_0,k_x,k_y,k_z)$ has two possible transverse polarization vectors \cite{Hagiwara:1985yu}:
\begin{eqnarray}
\epsilon^\mu_1(k) &=& \frac{1}{k_0 k_T}(0,k_x k_z,k_y k_z,-k_T^2 )\\
\epsilon^\mu_2(k) &=& \frac{1}{k_T}(0,-k_y,k_x ,0 ),
\end{eqnarray}
where $k_T=\sqrt{k_x^2+k_y^2}$. When $k_T=0$, the two polarization vectors become
\begin{eqnarray}
\epsilon^\mu_1(k) &=& (0,1,0,0 )\\
\epsilon^\mu_2(k) &=& (0,0,\frac{k_z}{\left|k_z\right|} ,0 ).
\end{eqnarray}
The circular polarization vectors of the photon is then defined by 
\begin{equation}
\epsilon^\mu_\pm(k) = \frac{1}{\sqrt 2}\left(\mp \epsilon^\mu_1-i\epsilon^\mu_2\right),
\end{equation}
where the two circular polarization vectors are related by complex conjugation,
\begin{equation}
\epsilon^\mu_\pm(k)  = e^{i \phi} {\epsilon^\mu_\mp(k) }^*,
\end{equation}
and 
\begin{equation}
\epsilon^\mu_\pm(\bar k)  = e^{i\bar \phi} \epsilon^\mu_\mp(k),
\end{equation}
and $\bar k^\mu=(k_0,-k_x,-k_y,-k_z)$. It should be noted that both phases depend on the phase convention for the definition of the polarization vectors. 

\subsection{Parity and CP violation}
From the last identity, one sees that the notion of circular polarisation is related to parity violation. Indeed, the sign of the spatial components of the four-momentum change under parity while the polarization vector stays invariant, i.e.
\begin{eqnarray}
k^\mu &\underrightarrow{P}& \bar k^\mu \\
\epsilon^\mu_\pm(k) &\underrightarrow{P}& \epsilon^\mu_\pm(\bar k) \propto\epsilon^\mu_\mp(k).
\label{kmu_parity}
\end{eqnarray}

\begin{figure}
\includegraphics[width=0.48\textwidth]{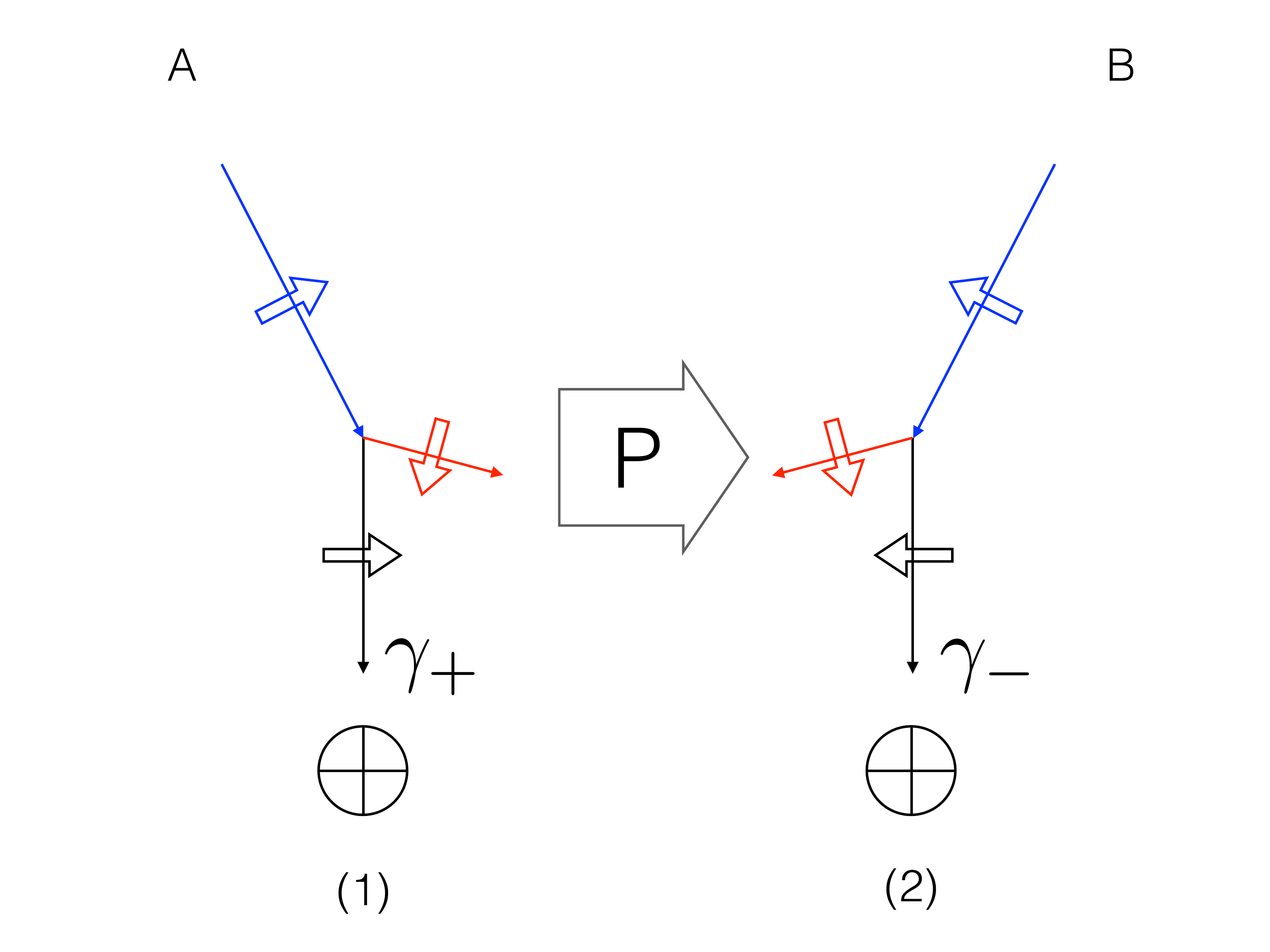}
\caption{Example of parity configuration for a decay. Regular and contour  arrows represent respectively the momentum and the spin or polarisation vector of each particle, as one should observe in the sky from Earth ($\bigoplus$)}
\label{fig:parity}
\end{figure}

A net circular polarisation is observed in the sky when there is an excess of one polarization state over the other.  Given  Eq.~\ref{kmu_parity}, 
this means that parity must be violated in at least one of the dominant photon emission processes. 

This is illustrated in Fig.~\ref{fig:parity}, where we represent the two-body decay of a "blue"  particle into a photon (black line) and a "red" particle for two parity-related configurations. Assuming that the "blue" particles are  homogeneously distributed in the sky (i.e. they are not all emitted in a particular direction), the two configurations 1 $\&$ 2 should produce the same number of photons with opposite polarisation, if parity is conserved. More specifically, if configuration (1) only produces photons with a positive circular polarisation, then configuration (2) should produce the same amount of photons with a negative circular polarisation, and vice versa (see e.g. Ref. \cite{Gabrielli:2005ek} for an analytic treatment in the case of meson and muon decay). 

Since the observed circular polarisation is determined by the difference between the number of  photon of each polarisation, there should be no observable circular polarisation in the sky for the configuration that we just described. If  parity is violated, then one should observe a net circular polarisation in the sky.  For example, if configuration 1 is the only process emitting a photon, then the net polarisation  will be positive (negative), depending on whether the photon has a positive (negative) polarisation.

Parity violation is not the only condition required to generate a net circular polarisation. If a decay or scattering process violates parity but is CP invariant, the rate for producing one photon in polarization state $\lambda$ from the initial state $i$ will be the same as the rate for producing a photon with the other polarization state ($-\lambda$) from the CP conjugate state $\bar{i}$. 
Alternatively, a process could violate P but conserve CP and it may still be at the origin of a net circular polarisation. This occurs if the number density of one of the particles in the initial state is not the same as the number density of its antiparticle. 
Therefore, two conditions are needed to generate a circular polarisation effect at the process level, namely:

\begin{enumerate}

\item \textbf{Parity must be violated}  

This typically occurs if one of the particles in the initial or final state is purely left or right-handed or if the mediator of the interaction process is more sensitive to one of the chiral components of the particles in the initial and final states. For example, the $Z$ boson interacts more with  left-handed  than  right-handed electrons. As a result a small circular polarisation is associated with M\o ller scattering, due to the exchange of a $Z$ boson.

\item \textbf{There must be an asymmetry in the number density of one of the particles in the initial state or CP must be violated.}
For example, the neutrino decay may give rise to  circularly polarised photons.  However the decay of antineutrinos of similar energy would produce the same amount of photons with opposite polarisation. This would eliminate the circular polarisation in the sky, unless the number densities of neutrinos and antineutrinos were not the same.  A difference in the number densities of neutrinos and antineutrinos is thus needed to generate a net circular polarisation signal in the sky.
\end{enumerate} 

Note that the only processes which can generate a circular polarisation are those where the initial state is not a CP eigenstate. For example, although  $e^+ e^- \rightarrow \nu \bar{\nu} \gamma$ violates parity, the fact that the initial state is a CP eigenstate prevents the generation of a circular polarisation at the process level. 
This condition excludes de facto any process involving ``symmetric" particles, such as Majorana particles in the initial state or two photons/gauge bosons in the final state.

\section{Polarisation as a new probe of the $\nu$ - $\bar{\nu}$ asymmetry \label{sec:neutrinos}}

\subsection{Neutrino decay}
The standard Model of particle physics predicts that the heaviest neutrinos radiatively decays into a lighter  neutrino plus a photon, through a 1-loop process \cite{Cowsik:1977vz,Raffelt:1988vd,Falk:1978kf}.  Since i) the photon can be emitted by either the  gauge boson $W^\pm$ or the charged lepton present in the loop and ii) the associated couplings are chiral (left-handed only) due to the presence of the W$^\pm$, such a decay is expected to generate a net circular polarisation of the photons thus emitted.

\subsubsection{SM neutrino decay rates and polarisation rates} 
The amplitude for the radiative decay of active neutrinos can be written as
\begin{equation}
\mathcal{A}(\nu_i \to \nu_j \gamma) = \sum_{l=e,\mu,\tau} U^\dagger_{il} \ U_{lj} \ A(m_l)
\end{equation}
where $U$ is the neutrino mixing matrix and $A(m_l)$ is the contribution from all the diagrams where there is a charged lepton $l$. 

In this equation, the dependence on the mixing factors out. Hence the amplitudes $A(m_l)$  only differ by the charged lepton mass. This is very important. Indeed, due to the unitarity condition, the amplitude $\mathcal{A}(\nu_i \to \nu_j \gamma) $ is expected to vanish if all the $A(m_l)$ are identical. Hence, the total amplitude is proportional to the charged lepton mass difference.

For purely Dirac neutrinos, the amplitude up to first non-vanishing order in $1/m_W^2$ \cite{Giunti:2014ixa,Pal:1981rm} is  given by 
\begin{equation}
\mathcal{A}(\nu_i \to \nu_j \gamma) = -i \sigma^{\mu\nu} \epsilon_\mu q_\nu (\mu_{ij}+i\varepsilon_{ij}\gamma_5),
\end{equation}
where $q$ and $\epsilon$ are respectively  the momentum and polarization of the photon and the terms $\mu_{ij}$ and 
$i \varepsilon_{ij}$ are
\begin{equation}
\begin{array}{c}
\mu_{ij}\\
i \varepsilon_{ij}
\end{array}
=
-\frac{3eG_f}{32\sqrt{2}\pi^2}\left(m_{\nu_i} \pm m_{\nu_j}\right)\sum_{l=e,\mu,\tau} U^\dagger_{il}U_{lj} \left(-2+\frac{m_l^2}{m_W^2}\right). \nonumber
\end{equation}

The constant term (i.e. the $-2$ term) is obviously independent of the mass of the charged lepton. Hence it  does not contribute to the amplitude when the mixing matrix is unitary. Therefore, the decay width of an active neutrino $i$ decaying into a lighter active neutrino $j$ is given by\footnote{If there is an heavier pure Dirac neutrino without the corresponding charged lepton, the above expressions are still valid but the first term in the expansion $m_l/m_W$ now contributes.
}
\begin{equation}
\Gamma(\nu_i \to \nu_j \gamma)=\frac{1}{8\pi}\left(\frac{m_{\nu_i}^2- m_{\nu_j}^2}{m_{\nu_i}}\right)^3\left(\left|\mu_{ij}\right|^2+\left|\varepsilon_{ij}\right|^2\right). 
\end{equation}

While the rate is expected to be strongly suppressed in the Standard Model, this radiative decay does lead to a net circular polarisation. Indeed, the decay rates into positive  and negative polarisations read as 
\begin{eqnarray}
\frac{\Gamma(\nu_i \to \nu_j \gamma_+)}{\Gamma(\nu_i \to \nu_j \gamma)}&=&\frac{\left|\mu_{ij}-i\varepsilon_{ij}\right|^2}{2\left(\left|\mu_{ij}\right|^2+\left|\varepsilon_{ij}\right|^2\right)}=\frac{m_{\nu_i}^2}{m_{\nu_j}^2+m_{\nu_i}^2} \label{decay_poli} \nonumber\\
\frac{\Gamma(\nu_i \to \nu_j \gamma_-)}{\Gamma(\nu_i \to \nu_j \gamma)}&=&\frac{\left|\mu_{ij}+i\varepsilon_{ij}\right|^2}{2\left(\left|\mu_{ij}\right|^2+\left|\varepsilon_{ij}\right|^2\right)}=\frac{m_{\nu_j}^2}{m_{\nu_j}^2+m_{\nu_i}^2}, \nonumber\\
\label{decay_polj}
\end{eqnarray}
respectively.  Note that these equations have been checked numerically against the loop-induced module of Madgraph5\_aMC@NLO \cite{Hirschi:2015iia} with the $R_2$ and UV vertices for the SM with massive Dirac neutrinos provided by NLOCT \cite{Degrande:2014vpa}.

In the presence of a strong mass hierarchy (for example $m_{\nu_i} > m_{\nu_j}$), the heaviest active neutrinos decay almost exclusively into $\gamma_+$ while, when the mass difference $m_{\nu_i} - m_{\nu_j}$ is small, one expects almost the same amount of $\gamma_+$ and $\gamma_-$ to be produced. 

Using these expressions, we estimate the decay width in the Standard Model to be \begin{equation}
\label{eq:decay_BSM}
\Gamma(\nu_3 \to \nu_1 \gamma) = 6.45\times10^{-52} \ \mathrm{s}^{-1}
\end{equation} 
and 
\begin{equation}
\label{eq:decay_BSM2}
\Gamma(\nu_3 \to \nu_2 \gamma) = 1.27\times10^{-51} \ \mathrm{s}^{-1},
\end{equation} 
thus leading to a large polarisation fraction 
\begin{equation}
\Gamma(\nu_3 \to \nu_1 \gamma_+)/\Gamma(\nu_3 \to \nu_1 \gamma)= 0.99
\end{equation}
and 
\begin{equation}
\Gamma(\nu_3 \to \nu_2 \gamma_+)/\Gamma(\nu_3 \to \nu_2 \gamma)=0.96,
\end{equation}
for  $m_{\nu_1}=8.5 \times 10^{-4}$ eV, $m_{\nu_2}=8.7 \times 10^{-3}$ eV, $m_{\nu_3}=5.016 \times 10^{-2}$ eV, $\theta_{12}= 0.588$, $\theta_{13}=0.154$, $\theta_{23}=0.78$, see \cite{Cuesta:2015iho} and assuming no CP phase.

Since antineutrinos produce the opposite polarisation, these results show that observing a net circular polarisation would simultaneously constrain the active neutrino decay rate, the neutrino mass hierarchy and the neutrino asymmetry.

\subsubsection{Polarised photon fluxes from neutrino decay in astrophysical sources} 

The flux of polarised photons ($\gamma_\pm$) produced by the radiative decay  of a population of neutrinos of mass  $m_{\nu}$ and energy $E_{\nu}$ in a source located at a distance $D$
can be written as 
\begin{eqnarray}
\phi_{\nu_i \rightarrow \nu_j \gamma_\pm} &=& \frac{\dot{n}_{\nu,0}}{4\pi D^2 } \, f_i \, \left( 1- e^{-\frac{m_{\nu_i}}{E_\nu} \frac{D}{c} \Gamma_{\nu_i \rightarrow \nu_j \gamma_\pm} } \right) \label{eq:SNnulife}\\
&\simeq&  \phi_{\nu,0} \ \frac{m_{\nu_i}}{E_\nu} \,  \frac{D}{c}   \, f_i \, \Gamma_{\nu_i \rightarrow \nu_j \gamma_\pm}, \label{eq:SNlongnulife} 
\end{eqnarray}
where $\Gamma_{\nu_i \rightarrow \nu_j \gamma_\pm}$ is the decay rate of neutrino mass eigenstate $i$, $f_i$  the fraction of this eigenstate produced by the source,   $\dot{n}_{\nu,0}$ the total  number of neutrinos produced per second by the source\footnote{The factor $m_\nu/E_\nu$ accounts for the Lorentz time dilation of the neutrino decay time with respect to its rest frame.} and 
$ \phi_{\nu,0} =  {\dot{n}_{\nu,0}}/{4\pi  D^2 }$ 
the neutrino flux emitted from the source decay. 

Since we expect a mixture of neutrinos and antineutrinos and, as mentioned in the previous section, antineutrinos should produce photons with opposite polarisation,  the net flux of polarised photons should be equal to 
\begin{equation}
\phi_{\gamma, pol} ^{\nu_i}= (\xi -\bar{\xi}) \ (\phi_{\nu_i \rightarrow \nu_j \gamma_+} - \phi_{\nu_i \rightarrow \nu_j \gamma_-}) \, , 
\end{equation}
where $\xi = n_{\nu_i}/(n_{\nu_i} + n_{\bar{\nu}_i})$ describes the proportion of neutrinos and  $\bar{\xi} = n_{\bar{\nu}_i}/(n_{\nu_i} + n_{\bar{\nu}_i})$ that of antineutrinos.  A negative flux $\phi_{\gamma, pol} ^{\nu_i}$ would indicate an excess of negative circular polarisation.
Note that here we assume that the decay rate of antineutrinos is the same as for neutrinos, which is the case if CPT is conserved. 
The flux of polarised photons from the decay of $\nu_i$  is therefore given by 

\begin{eqnarray} 
\label{eq:phi_cp_nu}
\phi_{\gamma, pol}^{\nu_i} &=& \phi_{\nu,0} \ \frac{m_{\nu_i}}{E_\nu} \,  \frac{D}{c}   \, f_i \,  \Delta_{CP } \ \Gamma_{\nu_i, pol}   
\end{eqnarray}
with 
\begin{equation}
\Delta_{CP } =   \left(\xi - \bar{\xi}\right)
\end{equation}
and 
\begin{equation}
\Gamma_{\nu_i, pol} =  \left( \Gamma_{\nu_i \rightarrow \nu_j \gamma_+} - \Gamma_{\nu_i \rightarrow \nu_j \gamma_-} \right) \, .
\end{equation}
In the case of a strong mass hierarchy, the decay only generates photons with a positive circular polarisation. Hence $\Gamma_{\nu_i, pol}  = \Gamma_{\nu_i \rightarrow \nu_j \gamma_+}, $ as we will consider in the remaining of this section.

Quantitative estimates of the polarised flux depend on the properties of the source. 

However, the inversion of Eq.~\ref{eq:phi_cp_nu} 
\begin{equation}
\Gamma_{\nu_i, pol} \simeq  10^{-11} \ \mathrm{s^{-1}} \ \left( \frac{D}{\mathrm{kpc}}\right)^{-1}  \frac{E_{\nu}}{m_{\nu_i}} \ (\Delta_{CP}  f_i)^{-1} \ \frac{\phi_{\gamma, pol}^{\nu_i}}{ \phi_{\nu_,0}} ,
\end{equation}
shows that requiring that the polarised flux be as large as the neutrino flux ($\phi_{\gamma, pol}^{\nu_i} \sim \phi_{\nu,0}$) implies a radiative decay rate that is forty orders of magnitude larger than the Standard Model predictions (see Eqs.~\ref{eq:decay_BSM} and \ref{eq:decay_BSM2}), if we also assume a source at a few kpc, $E_{\nu} > m_{\nu_i}$ and $\Delta_{CP}  f_i \sim 1$.

In reality,  
$\phi_{\gamma, pol}^{\nu_i}$ and $\phi_{\nu,0}$ are expected to be very different, but it is unlikely that their difference compensates the forty  orders of magnitude that stand at present. Hence, the circular polarisation signal from the  radiative  decay  of active neutrinos in the Standard Model should be very small; likely invisible in fact.

Neutrino lifetimes could nevertheless be much shorter than the SM's $\approx 10^{52}$ s, due to the presence of BSM physics (i.e. $\Gamma_{\nu_i, pol}$ could be much larger than the Standard Model value considered here). In this case, polarised radiative decay may lead to a visible polarised signature.

For example, in the case of a nearby supernova\footnote{Our most abundant neutrino source is the Sun, but it is too close to probe allowed lifetimes.}, located at 10 kpc from us and emitting a MeV neutrino flux  $\phi_{\nu,0} \sim 10^{10}$ cm$^{-2}$ s$^{-1}$. The expected photon flux for such a source is given by
\begin{equation}
\label{eq:phi_SN}
\phi_{\gamma,pol}^{\nu_i} \simeq 10^{16} f_i \,  \Delta_{CP } \ \left(\frac{m_{\nu_i}}{\mathrm{eV}}\right) \,\left(\frac{\Gamma_{\nu_i, pol}}{s^{-1}}\right) \ ph \, \mathrm{cm}^{-2} \, \mathrm{s}^{-1}.
\end{equation}

Recent simulations of core-collapse supernovae show that there is a dipolar asymmetry in neutrino lepton number emission, known as lepton-number emission self-sustained asymmetry (LESA) \cite{2014ApJ...792...96T}. Assuming that the neutrino mass eigenstates are produced  proportionally to the flavor eigenstate ratios, which are roughly $(\nu_e:\nu_\mu:\nu_\tau) \simeq (2:1:1)$ \cite{Janka:2016fox}, the normal hierarchy scenario predicts $f_i \sim 1/4$, while the inverted hierarchy (IH) predicts $f_i \sim 1/2$. Therefore, depending on the angle of the dipole with respect to our line of sight, LESA can yield asymmetries between $\nu_e$ and $\bar \nu_e$ resulting in $\Delta_{CP} \in [0.02,0.4]$ \cite{2014ApJ...792...96T}.  Putting these ingredients together in Eq.\ref{eq:phi_SN} and assuming $  m_{\nu_i} \gg m_{\nu_j}$, we obtain  a flux of circularly polarized $\sim$ MeV photons for the IH scenarios of:
\begin{equation}
\phi_{\gamma,pol}^{\nu_i} \simeq [1-20] \times 10^{14}  \left(\frac{m_{\nu_i}}{\mathrm{eV}}\right) \,\left(\frac{\Gamma_{\nu_i, pol}}{s^{-1}}\right) \ ph \, \mathrm{cm}^{-2} \, \mathrm{s}^{-1}. \nonumber
\end{equation}

The exact value of the flux then depends on the polarised radiative decay rate. The neutrino decay  cannot be too fast, i.e not before 
\begin{equation}
 t_{sun} > 10^6-10^9 \ \rm{s} 
\label{eq:sun}
\end{equation}
due to constraints from the Sun, see Ref.~\cite{Raffelt:1985rj}, but there is  an even stronger constraint. If the decay (and in particular the decay of the heaviest neutrino state) occurs after thermal decoupling ($T\sim$ MeV) and before recombination, the number of invisible relativistic degrees of freedom $N_{eff}$ will be depleted, which  will directly affect both estimates of the cosmological parameters and the shape of the CMB angular power spectrum.  Late decays may also distort the CMB spectrum due to the out-of-equilibrium energy injection. These cosmological considerations led  the authors of Ref. \cite{Ressell:1989rz} to the conservative bound $\Gamma_\nu^{-1} > t_{rec} \simeq 10^{13} \rm{s}.$
In reality, lifetimes on the order of the age of the Universe would still produce O(1) distortions in the CMB blackbody spectrum which have not been seen. Hence one can set the \textit{conservative} bound $\Gamma_\nu^{-1} > t_{0} =  4.3 \times 10^{17} $ \, s.

Such constraints thus allow a polarised gamma-ray flux from a 10 kpc Supernova (in the conditions we described) as large as 
\begin{equation}
\phi_{\gamma,pol}^{\nu_i} <  2 \times  10^{-2} \left(\frac{m_{\nu_i}}{\mathrm{eV}}\right) \ ph \, \mathrm{cm}^{-2} \, \mathrm{s}^{-1}. \label{eq:snflux}
\end{equation}
This value is much larger than the expected conventional flux at MeV for Cassiopeia A for example ($\Phi_{\rm{Cas \ A}} \sim 10^{-6} \rm{ph/cm^2/s}$) \cite{Saha:2014fga}, bearing in mind that weak processes in SN (such as $\nu \  e$ scattering, as discussed next or $\beta$ decay) provide competing mechanism for the generation of circular polarisation in these objects (though the resulting photons may be absorbed). 

\subsection{Neutrino-electron elastic scattering}

\begin{table*}[ht]
\begin{tabular}{l|c|c|c|c|c}
$\sqrt{S}$ & cross-section & $\gamma_+$ ($\%$) &$\gamma_-$  ($\%$) &net pol ($\%$) & $E_{\rm{min}}$ ($\gamma$)\\
\hline
1 TeV& 1.0 pb &  43 & 57 & -14.1 $\pm$ 1.2 & $100$ GeV\\
1 TeV & 3.1 pb &  47 & 53 & -6.3 $\pm$ 1.2 & $10$ GeV\\
1 TeV& 5.3 pb &  48 & 52 & -3.2 $\pm$ 1.2 & $1$ GeV\\
100 GeV& 0.35 pb &  36 & 64 & -28.5 $\pm$ 1.2 & $10$ GeV\\
100 GeV& 1.19 pb &  44 & 56 &-11.2 $\pm$ 1.2 & $1$ GeV\\
10 GeV& 0.0052 pb & 34 & 66 &-32.7 $\pm$ 1.2& $1$ GeV\\
10 GeV& 0.0185 pb & 44 & 56 & -12.6 $\pm$ 1.2 & $0.1$ GeV\\
\end{tabular}
\caption{Number of polarised photons produced  in  the process
$\nu_{\mu} e^- \rightarrow \nu_{\mu} e^- \gamma$, for a range of energies in the initial state  ($E_{CM}$) and different photon detection energy threshold ($E_{\gamma}$). The second column indicates the associated cross section, the third the  number of  $\gamma_+$, the fourth the number of $\gamma_-$, the fifth the degree of polarisation. The polarization for the conjugate process $e^+ \bar{\nu}_\mu \to e^+ \bar{\nu}_\mu \gamma$ are inverted. }
\label{tab:numu_e_scattering}
\end{table*}

\begin{table*}[ht]
\begin{tabular}{l|c|c|c|c|c}
$\sqrt{S}$&  Cross section &$\gamma_+$ (\%) &$\gamma_-$ (\%) &net pol (\%) & $E_{\rm{min}} (\gamma)$ \\
\hline 
1 TeV& 1.0 pb &  54 & 46 & 8.8 $\pm$ 1.2 & $100$ GeV\\
1 TeV & 3.0 pb &  52 & 48 & 4.2 $\pm$ 1.2 & $10$ GeV\\
1 TeV& 5.1 pb &  51 & 49 & 2.6 $\pm$ 1.2 & $1$ GeV\\
100 GeV & 0.30 pb & 55 & 45 & 9.8 $\pm$ 1.2 & $10$ GeV\\
100 GeV & 1.00 pb & 52 & 48 & 4.2 $\pm$ 1.2 & $1$ GeV\\
10 GeV& 0.0042 pb & 59 &41 & 17.2 $\pm$ 1.2 & $1$ GeV\\
10 GeV& 0.015 pb &  53 &47 & 6.4 $\pm$ 1.2 & $0.1$ GeV\\
\end{tabular}
\caption{Number of polarised photons in $e^- \bar{\nu}_\mu \to e^- \bar{\nu}_\mu \gamma$, for a range of energies in the initial state  ($E_{CM}$) and different photon detection energy threshold ($E_{\gamma}$). The second column indicates the associated cross section, the third the  number of  $\gamma_+$, the fourth the number of $\gamma_-$ and the fifth the degree of polarisation. The polarization for the conjugate process $e^+ \nu_\mu \to e^+ \nu_\mu \gamma$ are inverted.}
\label{tab:numubar_e_scattering}
\end{table*}

We now explore the possible generation of a circular polarisation due to neutrino scattering with electrons near astrophysical sources. 

Since $\nu e^- \rightarrow \nu e^- \gamma$ violates  P, a net circular polarisation is expected.  The  level of circular polarisation associated with this process  is more or less significant, depending on  the energy of the photon. Taking the specific case of $\nu_{\mu}$, we find that the level of polarisation grows with energy, and can reach  $ 25-30 \%$ for high energy photons, as shown in Table.~\ref{tab:numu_e_scattering}.  Although we only present the muon neutrino case,  $\nu \ e^-$ scattering gives a circular polarisation whatever the  neutrino flavour.
The cross section associated with this radiative process is slightly suppressed with respect to the elastic contribution $\nu_{\mu} e^- \rightarrow \nu_{\mu} e^-$. However it remains large  enough (a few pb) at high energies to promise a potentially visible signature.  
Table.~\ref{tab:numubar_e_scattering} shows the related process of antineutrino scattering with electrons $ \bar{\nu}_{\mu} e^- \rightarrow \bar{\nu}_{\mu} e^- \gamma$. 

As mentioned in the previous sections,  $ \bar{\nu} e^+ \rightarrow \bar{\nu} e^+ \gamma$ generates the opposite polarization to $\nu e^- \rightarrow \nu e^- \gamma$; and $ \bar{\nu} e^- \rightarrow \bar{\nu} e^- \gamma$  yields the opposite polarization to $ \nu e^+ \rightarrow \nu e^+ \gamma$.
Hence, the presence of positrons and antineutrinos in large quantities could suppress a potential circular polarisation signal. In most environments, the positron fraction is considerably suppressed with respect to the electron fraction. Hence we do not expect the positron contribution to significantly alter the circular polarisation signal from $\nu e^- \rightarrow \nu e^- \gamma$. However the presence of antineutrinos could suppress the signal since we do not know if there is a 
large asymmetry between the neutrino and antineutrino number densities, although by comparing Tables \ref{tab:numu_e_scattering} and \ref{tab:numubar_e_scattering} it is interesting to note that a net polarisation remains even in the case of equal $\nu$ and $\bar \nu$ densities.

The detection of a gamma-ray circular polarisation signal in the direction of an astrophysical neutrino  source could therefore be used to i) tag the presence of neutrino-electrons interactions,  ii) help recover the initial energy of the electron-neutrino pair in the initial state (depending on the level of polarisation) and iii) eventually reveal the existence and magnitude of a neutrino-antineutrino asymmetry near the source. 

All these results can of course be generalised to new BSM neutrino-electron interactions, bearing in mind the existence of stringent constraints (see Ref.~\cite{Cerdeno:2016sfi} for the latest summary).

\section{Dark Matter \label{sec:DM}} 
We now investigate whether a polarisation signal can be used to search for dark matter particles in the sky.

\subsection{Sterile neutrino decay}
Sterile neutrinos have been proposed in an attempt to explain neutrino masses through the see-saw mechanism. The particular case of keV sterile neutrinos has drawn considerable attention as these particles are perfect examples of Warm Dark Matter candidates in FLRW cosmology frameworks, where one assumes only one dark matter candidate. 

In left-right models (see Ref.~\cite{Mohapatra:1974hk, Mohapatra:1974gc}), sterile neutrinos couple to active neutrinos through a very small mixing angle $\theta$ which makes them potentially visible. Indeed, through this angle, they can then decay into an active neutrino and a monoenergetic photon (with energy $E = m_{\nu_s}/2$), which provides a unique signature  of their presence.   
As this radiative decay diagram is similar to the radiative neutrino decay\footnote{Thanks to a much larger mass, the decay width can be significantly larger than the active neutrino decay rate, despite the smaller mixing.},  we expect  the sterile neutrino decay to produce a circularly polarised line emission. However it is worth remembering that  the presence of anti-sterile neutrino would  give rise to photons with opposite polarisation and could therefore suppress the expected net circular polarisation from the sterile neutrino decay in the absence of  a number density asymmetry. 

Quite recently, tentative hints of a 3.5 keV line detection have been interpreted as possible evidence of sterile neutrinos decay. The sterile neutrino would then have a mass of $m_{\nu_s} \simeq 7$ keV. The first indication for this signal was obtained by Ref.~\cite{Bulbul:2014sua}  who analysed the XMM-NEWTON data for the  Perseus cluster, Coma$+$Centaurus$+$Ophiuchus clusters and 69 other clusters. Ref.~\cite{Bulbul:2014sua} also analysed the Chandra ACIS-S/I data of the Perseus cluster, confirming that this was not an instrumental effect and recovering evidence for  the line. Further evidence were gathered  by Ref.~\cite{Boyarsky:2014jta} who examined the XMM-NEWTON data of Andromeda and the  Perseus cluster.   

Despite these exciting claims, the presence of a 3.5 keV line has not been firmly established as yet. Indeed these results are in tension with the Suzaku data \cite{Urban:2014yda}, which show no indication of a 3.5 keV line in the Coma, Virgo and Ophiuchus clusters. More worrying perhaps, these findings are in tension with the Hitomi observations of the perseus cluster, which do not support an excess of 3.5 keV photons in this cluster \cite{Aharonian:2016gzq}.  

The physical interpretation of the line is actually as disputed as its presence. In addition to sterile neutrinos,  potential explanations range from a potassium origin \cite{bananas} to systematic error.  However, while the sterile neutrino/dark matter interpretation has been challenged by many authors  \cite{bananas,Boyarsky:2014paa}, it is not  in contradiction with observations of dwarf galaxies \cite{Malyshev:2014xqa} or the galactic centre \cite{Riemer-Sorensen:2014yda}, although analyses of stacked galaxy spectra using the Chandra and XMM data cast some doubt on the origin of the line \cite{Anderson:2014tza}.

The exact mass and lifetime of the sterile neutrino which is needed to  explain the signal varies from analysis to analysis, since  the flux depends on the exact set of observations that has been considered. However, 

\begin{itemize}
\item In \cite{Bulbul:2014sua}, the authors found a sterile neutrino mass of $m_{\nu_s} =  7.1 \pm 0.07$ keV and a mixing angle of $\sin^2 (2 \theta) \simeq 7 \times 10^{-11}$, using a stacked analysis of 73 galaxy clusters with redshift ranging from 0.01 to 0.35. 

\item In \cite{Boyarsky:2014jta}, the authors reported a sterile neutrino mass of $m_{\nu_s}= 7.06 \pm 0.06$ keV  and a mixing angle in the range $\sin^2(2\theta) = (2-20) \times 10^{-11}$ (taking the column density --- i.e. the integral of the DM density over the line-of-sight --- $S = 600 \ M_{\odot}/\rm{pc}^2$ from the Perseus cluster and Andromeda). 
\end{itemize}

Hence, if this hypothesis is correct and there is an asymmetry, one should also see a polarised monochromatic line in the direction of these clusters, especially since  the lifetime should be 15 orders of magnitude shorter than that of the active (SM) neutrinos.

Taking  $m_{\nu_s} =  7$ keV  as a benchmark point and a decay rate of 
\begin{equation}
\Gamma_{\nu_s} = 1.38 \times 10^{-30} \ \left(\frac{\sin^2 (2\theta)}{10^{-8}} \right) \ \left(\frac{m_{s}}{\rm{keV}} \right)^5 \ \rm{s^ {-1}},
\end{equation}
  \cite{Boyarsky:2014jta}\footnote{We note a factor 2 with respect to \cite{Loewenstein:2009cm}.}, 
we find that the flux of photons expected from the decay of sterile neutrinos in a cluster is given by 
\begin{equation}
\phi_{pol} = \Delta_{CP}^{\nu_s} \ \Gamma_{\nu_s} \ \int_{FoV} d\Omega \int ds(r) \  \frac{\rho_{\rm DM } (r)}{m_{\rm DM}} (1+z)^{-1} \ ,  
\end{equation}
where $\Delta_{CP}^{\nu_s}$ measures the sterile neutrino-antineutrino asymmetry (assuming that the decay rates of the sterile neutrino and antineutrino are equal and that the decay gives rise to only one photon polarisation). 
The integrals of the DM profile $\rho_{\rm DM } $ are over the line-of-sight  and the angular field-of-view (FoV) of the experiment.  

The radial dependence of DM profiles in a wide class of cosmological structures is not known at small scales. Depending on the size of the objects and the baryonic physics, the profiles tend to be either cuspy or cored. This uncertainty (and the steepness of the profile in the inner part of the cluster) is generally averaged out once one takes into account the angular resolution of the experiment, but one needs to specify the profile nonetheless to make the theoretical prediction.

Focusing  on the Perseus cluster and the signal from the core ($z=0.0179$, $r_s =  164 \ h^{-1} \ \rm{kpc}$, $d=70$ Mpc), where analysis of the XMM-NEWTON and Chandra data seem to indicate an excess (unlike Hitomi's), one should then find a flux of polarised 3.5 keV photons of 
\begin{equation}
\phi^{Perseus}_{ pol, \ 3.5 keV} \sim  \ 10^{-6} \ \Delta_{CP}^{\nu_s} \ \  ph \ \rm{cm^{-2} s^{-1}},
\end{equation}
(depending on the exact FoV, here assumed to be about 1 arcmin$^2$) for a mixing angle of $\sin^2 (2 \theta) \sim 7 \times 10^{-11}$ (corresponding to a decay rate of $\Gamma_{\nu_s} \simeq 1.63 \ 10^{-28} \rm{s^{-1}}$), as suggested in Ref.~\cite{Franse:2016dln}.

\subsection{DM  scattering off SM particles}

DM could also be scattering off charged (SM) particles, especially in DM halos where there is a large number of electrons (e.g. SN) and  protons emitters (e.g. AGNs).  If the interactions occur through a mediator that does not couple equally to left and right SM components, then one expects a net circular polarisation to emerge from this scattering process from parity violation.

\subsubsection{ Neutralinos scattering off  electrons }

As an example, we consider the case of neutralinos in the galactic halo  scattering off high energy electrons  \cite{Gorchtein:2010xa}.  
Neutralinos have been the leading DM candidate for several decades. Their indirect detection signatures (mostly annihilation) have been thoroughly studied, but the associated polarisation signal has not been computed.

\begin{figure*}
\includegraphics[width=0.70\textwidth]{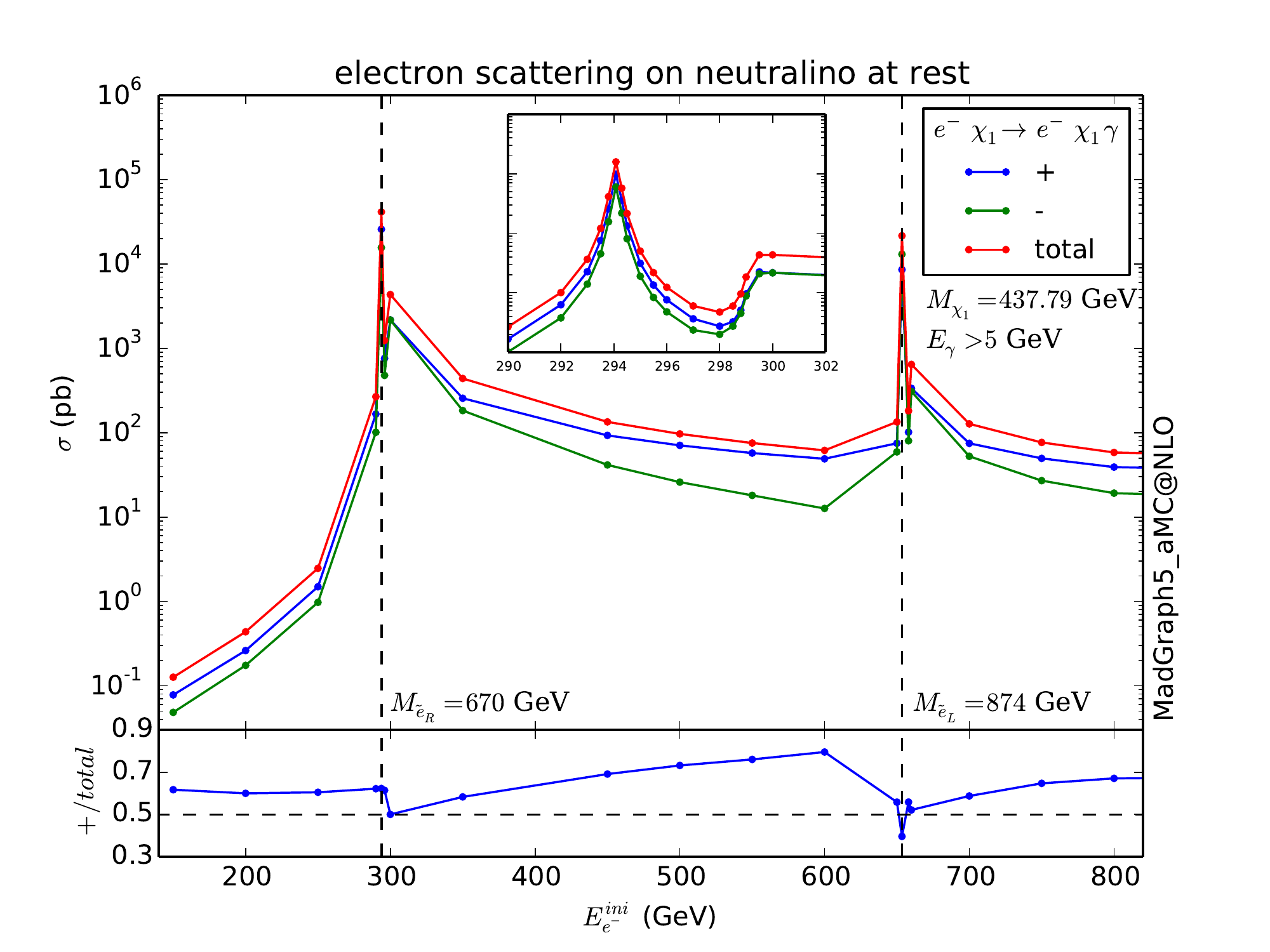}
 \caption{Cross-section and polarization fraction for electron scattering on the lightest neutralino as a function of the energy of the electron, with a cut-off on the energy of the photon at 5 GeV. Computations were done using the Madgraph5\_aMC@NLO package~\cite{Alwall:2014hca}.}
 \label{fig:en1cspol}
\end{figure*}

If they constitute a large fraction of the DM halo, then their interactions with ambient cosmic rays  (e.g. $\chi \ e^- \rightarrow \chi \ e^- \ \gamma$, $\chi \ e^- \rightarrow \chi^- \ \nu \ \gamma$,  $\chi \ p \rightarrow \chi \ p \ \gamma$) near accelerator sites can produce circularly polarised gamma-rays.  Focusing on the elastic scattering process, and assuming that neutralinos are at rest (which is a valid hypothesis given the DM virial velocity in the halo), we find 
that $\chi e^- \rightarrow \chi e^- \gamma$   gives rise to a large circular polarisation signal  through the $Z$ boson and (left and right) selectrons exchange, as shown in Fig.~\ref{fig:en1cspol}.

To illustrate the effect,  we have imposed a minimum energy detection threshold for the photon of $E_\gamma>5$ GeV, to avoid the tree level IR divergence\footnote{The electron mass is also non-zero.} and used the CMSSM benchmark point (best fit scenario) of Refs.~\cite{Beskidt:2014kon,Beskidt:2014oea} as this point satisfies LHC, flavour and DM constraints,  with 
\begin{eqnarray}
sign(\mu)&>&0 \nonumber\\
m_0&=&550 \ \rm{\, GeV}  \nonumber\\
m_{1/2}&=&1020 \rm{\, GeV} \nonumber\\
\tan\beta&=&19.16 \nonumber\\
A_0&=&-2878 \rm{\, GeV}. \nonumber
\end{eqnarray}
We note that the full LHA spectrum was obtained with softsusy-3.7.3 \cite{Allanach:2001kg} and only differs by about 1\% from the one in \cite{Beskidt:2014kon,Beskidt:2014oea}.

In the scenario considered above, the neutralino mass is 437.79 GeV.  The two peaks seen in  Fig.~\ref{fig:en1cspol} are due to the resonant production of  right ($M_{e_r}=670$ GeV) and left ($M_{e_l}=874$ GeV) selectrons,  decaying into $\chi e^- \gamma$. The dip after each resonance is due to the fact there is not enough energy to produce photons with energy greater than 5 GeV at the resonance, suppressing the production cross section. 

\begin{figure*}
\includegraphics[width=0.48\textwidth]{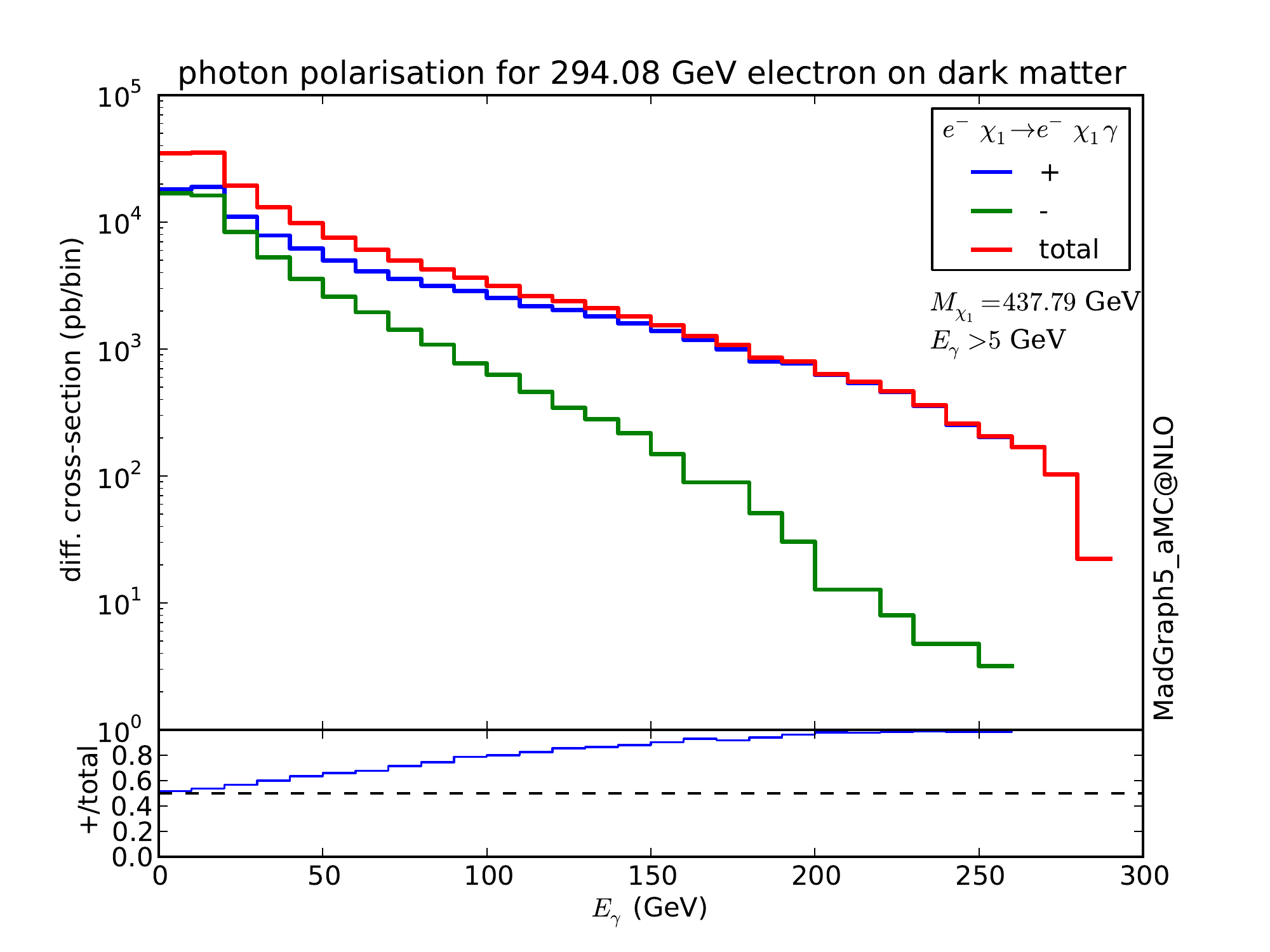}
\includegraphics[width=0.48\textwidth]{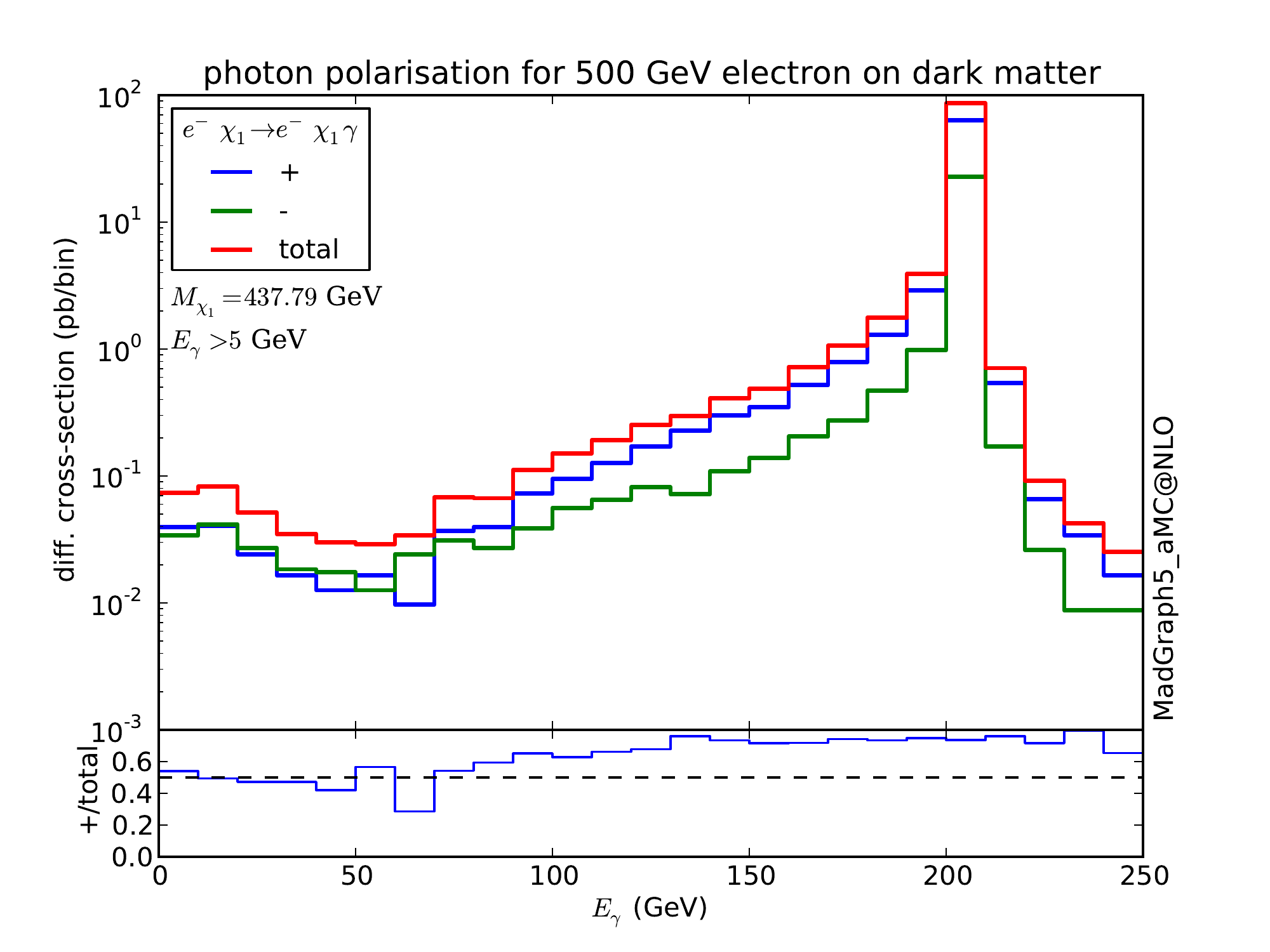}
\includegraphics[width=0.48\textwidth]{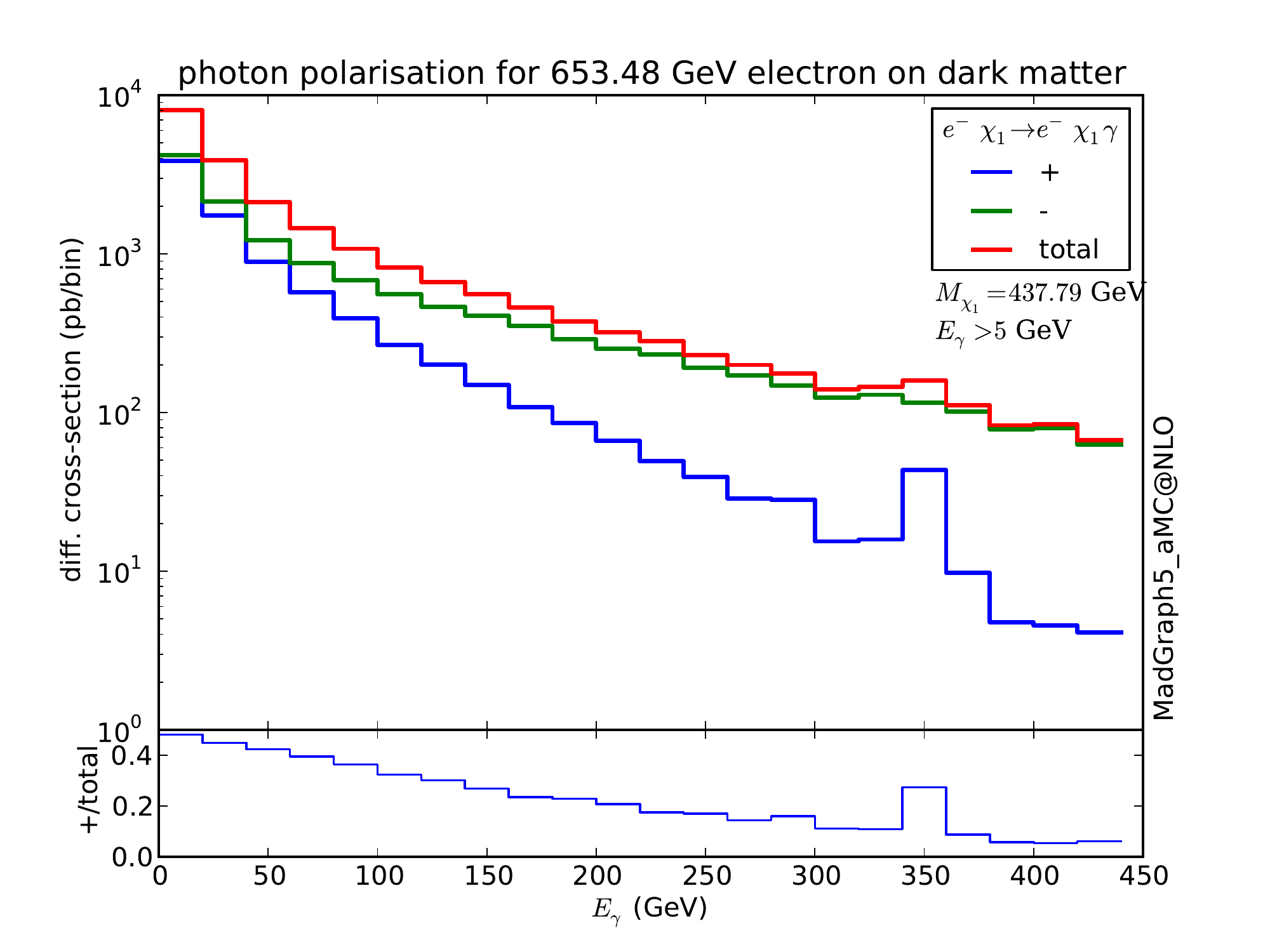}
\caption{Photon energy spectrum and polarisation fraction associated with the radiative  neutralino-electron elastic scattering process as a function of the photon energy  for three different energies of the incoming electron.  The energies for the tirst and third plots have been chosen so as to fall on the right and left selectron resonances respectively}
\label{fig:en1spectrum}
\end{figure*}

The photon spectra resulting from the $\chi e^- \rightarrow \chi e^- \gamma$ elastic scattering collisions are shown in Fig~\ref{fig:en1spectrum} for three electron energies (294.08, 500, 653.48) GeV.  For each of them, we show the total spectrum, corresponding to the sum of the two polarisation components (red curve), the spectrum associated with the positive circular polarisation (blue curve) and that associated with negative circular polarisation (green curve). 

The polarisation fraction is shown for each case (see lower panel associated with each plot). The dashed line corresponds to 50$\%$ polarisation.  Any departure from this dashed line (i.e. whether the fraction is above or below) indicates the existence of a net circular polarisation (positive or negative). 

It is apparent from these plots that the polarisation fraction is  more pronounced for high energy photons than for soft photons. We see also that its exact value depends on which selectron contribution is dominant. Hence the polarisation fraction is strongly dependent on the electron energy (like is the photon emission spectrum in fact).

Of course, a net polarisation effect may not exist  if the source emit as many positrons as electrons since   $\chi_0 e^- \rightarrow \chi_0 e^- \gamma $ and $\chi_0 e^+ \rightarrow \chi_0 e^+$ leads to a circular polarisation of opposite sign, and with the same polarisation fraction. However such symmetry is unlikely. Indeed, a very significant electron-positron asymmetry has been measured locally by Pamela, AMS02 and indirectly by Fermi-LAT, on a wide range of energies \cite{Bazo:2016nkn,Adriani:2013uda,Grasso:2009ma}. Therefore the number of positrons at the source location should be smaller than that of electrons and a net circular polarisation effect from  $\chi_0 e^- \rightarrow \chi_0 e^- \gamma $ should remain. 

In any case, the observation of such a circular polarisation near an accelerator site could be used to constrain both the nature of dark matter and a possible electron-positron asymmetry  at the source location.  Because it carries more information than just the photon spectrum, circular polarisation  represents a new way to probe the DM interactions with SM particles.

\subsubsection{ A new way to probe the distribution of low energy positrons? }

In the previous section, we focused on neutralino-electron elastic scattering. However the same conclusion applies to generic DM candidates (and could be extended to inelastic scattering) since the model dependence  only arises through the mediator, the DM mass and their couplings to SM particles.  Here we entertain the idea that a circular polarisation signal could be used to tag the presence of low energy positrons in the dark matter halo and understand their origin. 

Indeed, observations of a 511 keV emission line from the Galactic Centre (GC) has been reported for more than five decades \cite{Dermer:1997cq,Jean:2003ci} but its origin is unknown.  While it has been established that the line originates from positronium formation and is therefore a manifestation of an anomalously large number of low energy positrons in the galaxy and in particular in the center of the galaxy,  their origin is still very mysterious.  

It is actually plausible that several sources, distributed over the galaxy, seeded these low energy positrons. Indeed recent analysis suggested that the emission could be due to the superposition of  i) a narrow bulge, ii) a broad bulge, iii) a central source and  iv) disc components \cite{Siegert:2015knp}.  

The  narrow bulge component is the brightest of all, and the most difficult source to explain as a result.  It is unlikely that conventional astrophysical sources be the explanation of this component emission since they would also contribute to the galactic disc and make it much brighter than it is \cite{Prantzos:2010wi}. An annihilating dark matter interpretation (as initially proposed in Ref.~\cite{Boehm:2003bt}) also appears rather unlikely \cite{Wilkinson:2016gsy}, despite a suggestive morphology that agrees well with the dark matter explanation  \cite{Siegert:2015knp} (though one may need to revisit the constraints from \cite{Ascasibar:2005rw} since the analysis of more recent data shows a somewhat different morphology). Radioactive sources might be an explanation \cite{Crocker:2016zzt}, although this would suggest that the source responsible for the narrow bulge emission is absent from the disc.

The positron annihilation rate is proportional to $n_{e^+}n_{e^-}$, and thus relies on a high electron density. Since the low energy positrons are essentially bathing in a dark matter halo, their scattering off the DM  could give rise to a circular polarisation signal (depending on the precise nature of the DM). As the positron fraction $f_p = n_{e^+}/(n_{e^+}+n_{e^-})$ rises, the polarized signal would get smaller than what is observed from scattering with ambient electrons only. Crucially, if there are regions of the galaxy where $f_p$ becomes larger than 0.5, a flip in the observed polarization would be produced, allowing regions of high positron production to be singled out. The study of gas-poor candidates such as Reticulum II, where a  hint of 511 keV emission was recently found \cite{Siegert:2016ijv}, would allow for an independent measurement of $e^+$ sources.  Finally we observe that the positronium state itself could give rise to a circular polarisation \cite{Pokraka:2016jgy} though it may be far too suppressed to be observed.

\subsubsection{Neutralino-neutrino interactions} 

Since DM is supposed to co-exist with ambient (stable) cosmic rays, it may also interact with one of the most abundant species around, namely neutrinos (depending on its nature). 
For very energetic neutrinos, these interactions can produce heavier BSM states, whose decays potentially produce polarised photons. 
For example, in the case of neutralino DM,   there could be processes  such as e.g. $\chi_0 \nu_e \to \chi^\pm W^\mp \gamma $ or  $\chi_0 \nu_i \to \tilde{f}^\pm_j  W^\mp \gamma $  which could generate a polarised signal. Focusing on $\chi_0 \ \nu_i \to \tilde{f}^\pm_j \ W^\mp  \ \gamma $, we note that this process could immediately lead to the final state
$$
\chi_0 \ \nu_i  \to \chi_0 \ \nu_j \ f^{\pm}_{SM, i}  \ f'^{\mp}_{SM, j}  \ \gamma,  
$$
where $f^{\pm}_{SM, i}$ and $f'^{\mp}_{SM, j}$ are two light SM fermions that can be produced on-shell.

There are 412 tree level Feynman diagrams that contribute to the $\chi_0 \nu_e \to \chi_0 \nu_\tau e^- \tau^+ \gamma$ process. 
However, for the benchmark point considered here,  the dominant contributions are from the diagrams where staus and tau sneutrinos are exchanged, as these are the lightest supersymmetric particles ($m_{\tilde{\tau}_{1,2}} = $  441, 800 GeV and $m_{\tilde{\nu}_{\tau}} = $ 796 GeV).  Diagrams involving sleptons, the lightest chargino and the next to lightest neutralino  also contribute to this process but are ten times smaller. All the other diagrams are essentially negligible as they involve particles heavier than the TeV  scale. 

\begin{figure}
 \includegraphics[width=0.48\textwidth]{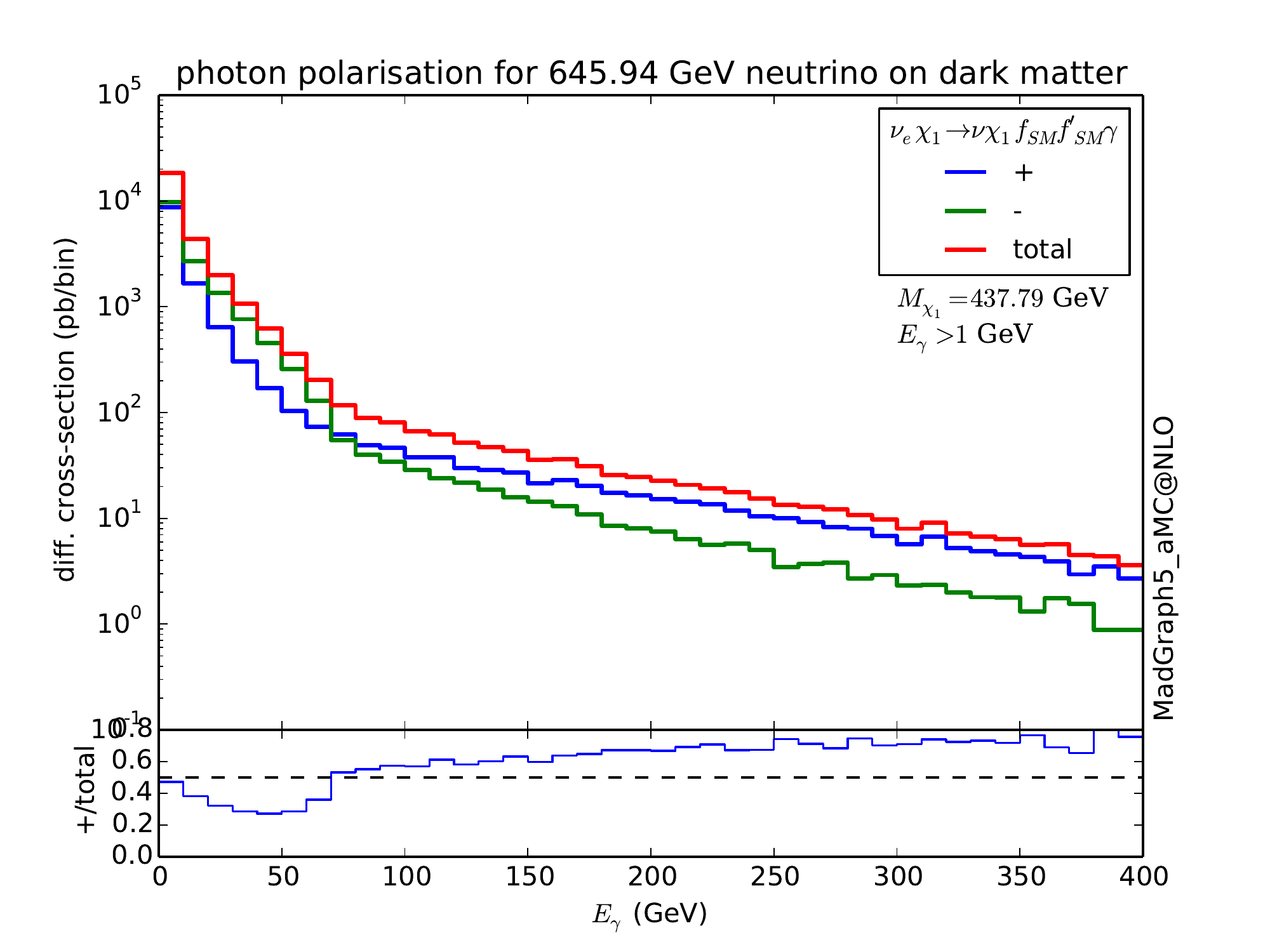}
 \caption{Photon spectrum for electron neutrino scattering at the sneutrino resonance ($M_{\tilde{\nu_e}}=870$GeV) with a cut on the photon energy at 1 GeV}
 \label{fig:ven1}
\end{figure}

The presence of a net circular polarisation is shown in Fig~\ref{fig:ven1},  where one can see that the fraction and sign of the polarisation oscillate depending on the photon energy. 
 Various contributions are actually responsible for the change of polarisation around 70 GeV. However, when the tau sneutrino is produced on-shell, the photon maximum energy is around 70 GeV and the dominant polarisation is negative. This subprocess contributes to more than half of the cross-section. 

\section{Conclusion \label{sec:conclusion}}

In this paper we have presented several examples of particle physics processes that could induce a net circular polarisation in the gamma ray sky. We have shown that given the known parity-violating processes in the Standard Model, the observation of polarized gamma rays can be used to reveal or constrain CP asymmetries in the neutrino and dark matter sectors.

We showed that large net circular polarisation could arise from the decay of active or sterile neutrinos, or from DM scattering with Standard Model particles. 

Measuring a signal in the direction of galaxy clusters could demonstrate the existence of keV-scale sterile neutrino and a sterile neutrino and antineutrino asymmetry. 

We also showed that processes such as the scattering of neutrinos or DM with electrons and neutrinos could potentially lead to a circular polarisation.  Measuring this effect  not only would single out the nature of the DM and the energy of the cosmic rays but it would also enable to measure an asymmetry in the number density of dark matter (for DM-SM scattering) and neutrinos (for $\nu$--SM scattering). 

More work is needed to determine whether these signatures can lead to large enough signals to dominate potential  circular polarisation emission from astrophysical background sources. 
However, our work shows that circular polarisation measurements can be used to gather a wealth of information about the particle physics processes in the Universe and the distribution of particles (including dark matter, neutrinos, electrons and positrons). Hence we conclude that the new generation of X-ray and gamma-ray polarimeters could be essential experiments to help us progress in our understanding of the fundamental constituents of the Universe.

\section*{Acknowledgements}
We thank Ryan Wilkinson with whom CB started the work on sterile neutrino polarisation, as well as A. Ibarra, P. Jean, T. Lacroix for illuminating discussions. 
OM would like to thanks the CERN TH division for its hospitality.
ACV is supported by an Imperial College London Junior Research Fellowship. CB was supported by the DFG cluster of excellence ʻOrigin and Structure of the Universe (www.universe-cluster.de). 
OM is a fellow of the Belgian Pole d'attraction Inter-Universitaire (PAI P7/37).

\section{Appendix }

\subsection{Some no-go cases \label{failedcases}}

Some processes may appear as promising candidates to generate circular polarisation but do not eventually lead to an observable effect.  Here we mention a few which have a pedagogical value. 
This is true in particular for neutralino pair annihilations.  One of the most important diagrams for the computation of the neutralino relic density is the neutralino 
pair annihilation into $e^+ e^-$ through the exchange of a slepton. Owing to the differences in selectron left ($\tilde{e}_L$) and selectron right  ($\tilde{e}_R$) masses, one may expect that $\chi \ \chi \ \rightarrow \ e^+ \ e^- \  \gamma$ could produce a, small, net circular polarisation. However 
the  CP  conjugation of this process leads to the exact same initial and final states. Hence there is no net circular polarisation to be expected from these processes. Other final states (such as $W^{\pm} H^{\mp}$) also individually  lead to large polarisation effects (about 25 $\%$ in the case of $W^{+} H^{-}$). However, since the $W^{-} H^{+}$ final state also gives the same amount of opposite polarisation, there is no net/overall circular polarisation out of these processes. Similarly sneutrino pair annihilations into $H^+ W^- \gamma$ lead to photons of opposite polarisation as sneutrino pair annihilations into  $H^- W^+ \gamma$. Hence no net circular polarisation is to be expected in this case.  Other sneutrino pair annihilation processes, invariant under parity conjugation, do not lead to a circular polarisation signal either.

\bibliography{darkpol.bib}

\end{document}